\documentclass[conference]{IEEEtran}

\usepackage{amssymb}
\usepackage{amsmath}

\usepackage[colorlinks=true, linkcolor=blue, citecolor=blue, urlcolor=blue]{hyperref}

\usepackage{graphicx}
\usepackage{booktabs}
\usepackage{multirow}
\usepackage{url}

\usepackage{fancyhdr} 
\fancypagestyle{firststyle}
{
   \fancyhf{}
   \fancyfoot[C]{\scriptsize{\textit{Proceedings of the Eighth International Conference on eDemocracy \& eGovernment (ICEDEG 2021)}, Quito (online), pp.~205--209, IEEE. This is the author's copy. The publisher's copy is available online at \url{https://dx.doi.org/10.1109/ICEDEG52154.2021.9530996}.}}
}

\begin{document}

\title{Assessing the Readability of Policy Documents on the Digital Single
  Market of the European Union}
\author{
\IEEEauthorblockN{Jukka Ruohonen}
\IEEEauthorblockA{University of Turku, Finland \\
Email: juanruo@utu.fi}
}

\maketitle

\begin{abstract}
Today, literature skills are necessary. Engineering and other technical
professions are not an exception from this requirement. Traditionally, technical
reading and writing have been framed with a limited scope, containing
documentation, specifications, standards, and related text types. Nowadays,
however, the scope covers also other text types, including legal, policy, and
related documents. Given this motivation, this paper evaluates the readability
of 201 legislations and related policy documents in the European Union (EU). The
digital single market (DSM) provides the context. Five classical readability
indices provide the methods; these are quantitative measures of a text's
readability. The empirical results indicate that (i) generally a Ph.D. level
education is required to comprehend the DSM laws and policy documents. Although
(ii) the results vary across the five indices used, (iii)~readability has
slightly improved over time.
\end{abstract}

\begin{IEEEkeywords}
readability index, comprehension, literature skills, text mining, legal
texts, law, digital single market, EU
\end{IEEEkeywords}

\section{Introduction}

\thispagestyle{firststyle} 

Comprehension of different texts is a requirement of today's life. No matter of
their background, citizens need to understand many text types to participate in
a society and manage their lives. At the same time, surveys around the world
have reported declines in literature skills, generally defined not only as the
ability to read and write but to also understand, evaluate, and use written
texts~\cite{OECD19}. There are many reasons behind the declines. Technology is
one factor. Although existing results are not definite~\cite{Verheijen20}, smart
phones and social media are among the technological factors partially explaining
the declines. When the declines are coupled with other trends, such as today's
avalanches of misinformation and disinformation, the societal consequences may
be~ruinous in the long-run.

Technical professions do not exempt from literature skills. Programmers need to
write documentation and understand the documentation of other
programmers. Engineers need to read and write technical specifications. Thus,
language and literature skills, critical thinking, and related ``humanistic''
skills have long been on the agenda in engineering education
programs~\cite{Abdulwahed13}. There are even online courses specifically for
technical writing skills~\cite{Google21a}. At the same time, the skill
requirements have likely increased and expanded in their scope.

Nowadays, engineers need to comprehend also legal documents, particularly in
case legal counseling is not attainable, as is often the case in start-ups and
small companies. To this and other ends, educators have long tried to expose
students to different text types~\cite{Conley11}. To some extent, the attempts
have beared fruit. For instance, anecdotal evidence hints that engineering and
computer science students are receptive to the grammar and vocabulary of
law~\cite{Hildebrandt20}. Yet questions remain whether and how well they are
able to understand and translate law into technical specifications and
implementations. In practice, collaboration between lawyers and engineers is
often required for such translations~\cite{Hjerppe19RE}. To ease the
translations, different formalization methods and tools have long been
proposed~\cite{Mommers09}. Eventually, such technical solutions may eliminate
the need to have lawyers in situations requiring the engineering of requirements
from law. These requirements have increased particularly in the EU as regulation
of the information technology sector has tightened; therefore, the DSM laws and
policy documents provide an ideal case study. But engineers and other technical
professions are obviously not the only ones required to read, interpret, and
understand law and its requirements.

The comprehension of law by ordinary citizens remains the most pressing
issue. And existing results are not encouraging. For instance, even many legal
counseling websites have been observed to be beyond the comprehension of those
with weak literacy skills~\cite{Dyson16}. The problem can be argued to contain
two parts. The first is intentional obfuscation. And, indeed, it is not
difficult to find arguments that companies and their legal representatives
deliberately obfuscate documents in order to evade transparency
requirements~\cite{Rutherford03}. Nor is public sector administration immune to
such arguments. The second part derives from the ways lawyers, academics, and
other professionals, including legislators, communicate in writing. The topics
covered by them are often complex and hence the writing tends to be
complex. Sometimes, though, complexity only serves complexity. Lawyers, for
instance, have long been accused of writing gobbledygook~\cite{Frooman73}. Other
professionals, including civil servants, are often no better in this
regard. Therefore, simple and understandable writing has long been seen as a
part of good administration. The same applies to law-making for which language
improvements have long been recommended~\cite{Karpen08}. This point motivates to
ask a research question (RQ$_1$): \textit{how readable are the DSM-specific laws
  and policy documents?} The framing to the DSM can be justified with the
earlier remarks on technical professions; these are the laws and policy
documents engineers specifically should be able to understand. The second
research question (RQ$_2$) is about validity: \textit{are there statistical
  differences between five typical readability indices?} The final RQ$_3$ is
longitudinal: \textit{has the readability of the laws and documents improved
  over~time}?

\section{Related Work}

The paper shares a long tradition of related work. Readability first became a
research topic already in the 1920s. Ever since, different quantitative indices
have been proposed to gauge the perceived readability of a text. The ideal has
been simplicity. Plain language, plain prose, or plain English have been the
terms used to describe this ideal. It has been endorsed by authors of both fact
and fiction. Actually, the classical assessments and opinions of both author
types have been highly similar, as becomes evident by comparing the 1946 works
of Flesch~\cite{Flesch60} and Orwell~\cite{Orwell46}, respectively. The former
author also developed a quantitative index for his ideal.

From the early 1970s onward, the Flesch's classical readability index was met by
multiple competing indices. Yet these competing indices never abandoned the
ideal of simplicity. Simple is usually better when a topic is
complex. Therefore, a classical application domain has been technical writing by
engineers \cite{McClure87, Zhou17a}. But the indices have been also applied for
many other purposes. Other application domains include the evaluation of
financial reports~\cite{Sun14}, websites~\cite{Leong02} and their privacy
policies~\cite{Fabian17}, tweets by politicians who speak at the level of
4--5:th graders~\cite{Kayam17}, fake online reviews~\cite{Wang18a}, and consent
forms for scientific research~\cite{Santele19}, to name some~examples.

Some previous work exists also for using indices to evaluate the readability
laws and related legal documents. The results have not been surprising. The
comprehension requirements have been observed to be beyond the educational
attainment of most people---and particularly of those who would most need
information about their rights~\cite{Arkell78, Tan92}. Even the instructions
delivered to juries in common law legal systems have been measured to be beyond
the literature skills of many adults~\cite{Small13}. The decisions reached by
some courts have also become more difficult to read over time, which, as such,
reflects the increasing complexity of many legal
questions~\cite{Whalen15}. However, no previous research seems to exists in the
EU context according to a reasonable literature search. Regarding the
engineering context, it is also worth remarking the work that has focused on the
understandability of the general logic of law~\cite{Ploch93}. But it is
difficult to understand a logic when a text describing the logic is difficult to
understand. This point provides a justification for evaluating the overall
readability of legislations in the EU.

\section{Data}

The dataset covers the main documents on the digital single market of the
EU. Although there exists no single database that would cover everything about
the DSM, the EU has still provided a portal that provides summaries about key
topical policy areas. The area reserved for the DSM was used to assemble the
dataset by covering all documents except archival material. Based on this
portal~\cite{EU21b}, five domains of the DSM are present: general rules,
electronic communication networks, personal data and privacy, copyright and
audiovisual material, and data economy and data protection. The data collection
resembled the so-called snowball sampling~\cite{Biernacki81}: for each domain,
all hyperlinks were visited, and the documents mentioned on the web pages were
collected as long as these were directives, regulations, or communications
(COM), staff working documents (SWD), recommendations, decisions, or joint
declarations (JOIN) of the European Commission or the Council. Hence, references
to informal documents, treaties, corrigenda, court cases, decisions of
non-legislative EU institutions, and related document types were excluded. Both
current and deprecated laws were included in the dataset but only insofar as
these were explicitly linked on the web pages.

\begin{figure*}[p!]
\centering
\includegraphics[width=\linewidth, height=17.5cm]{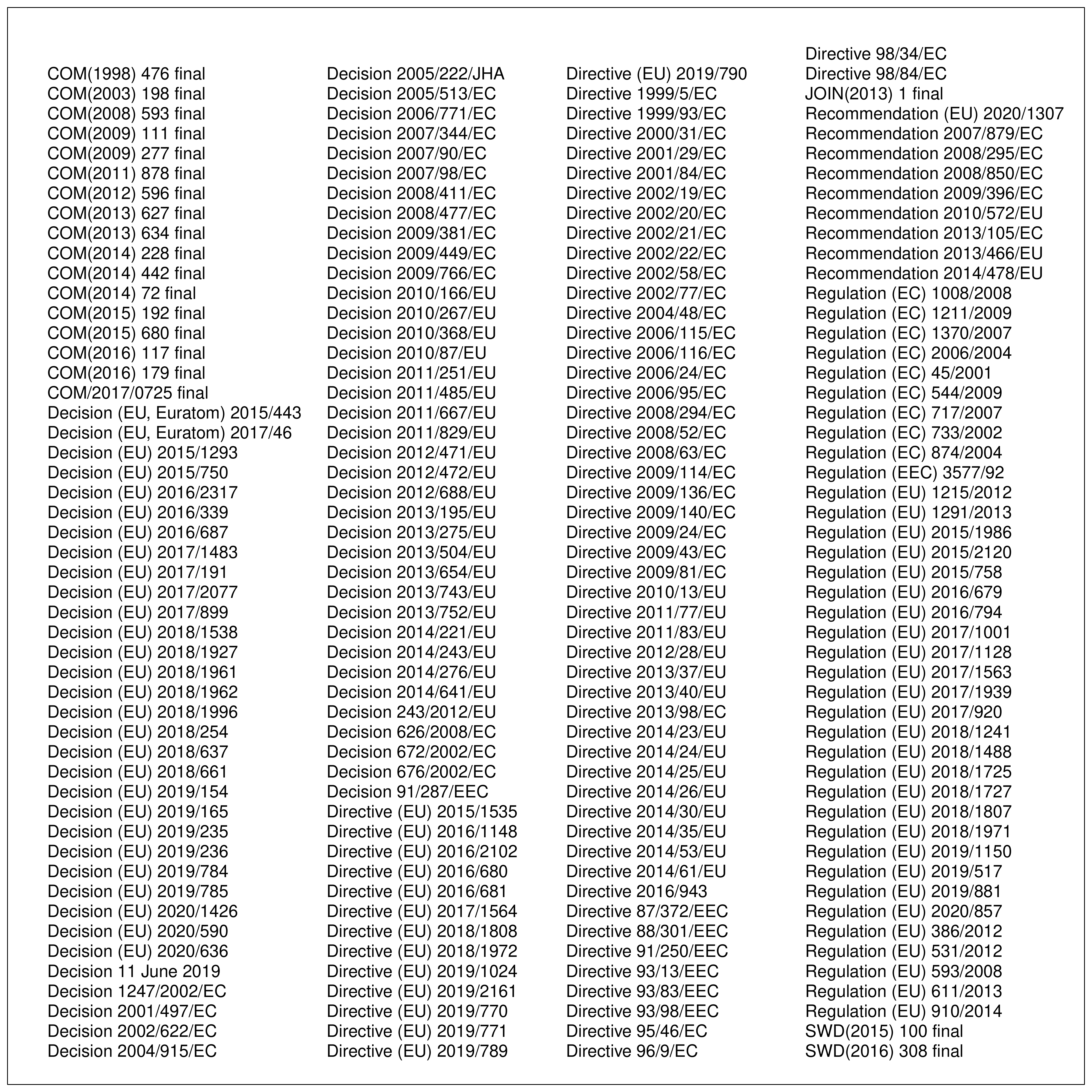}
\caption{Legal and Policy Documents Included in the Dataset ($n = 201$)}
\label{fig: regs}
%
\vspace{20pt}
%
\centering
\includegraphics[width=\linewidth, height=4cm]{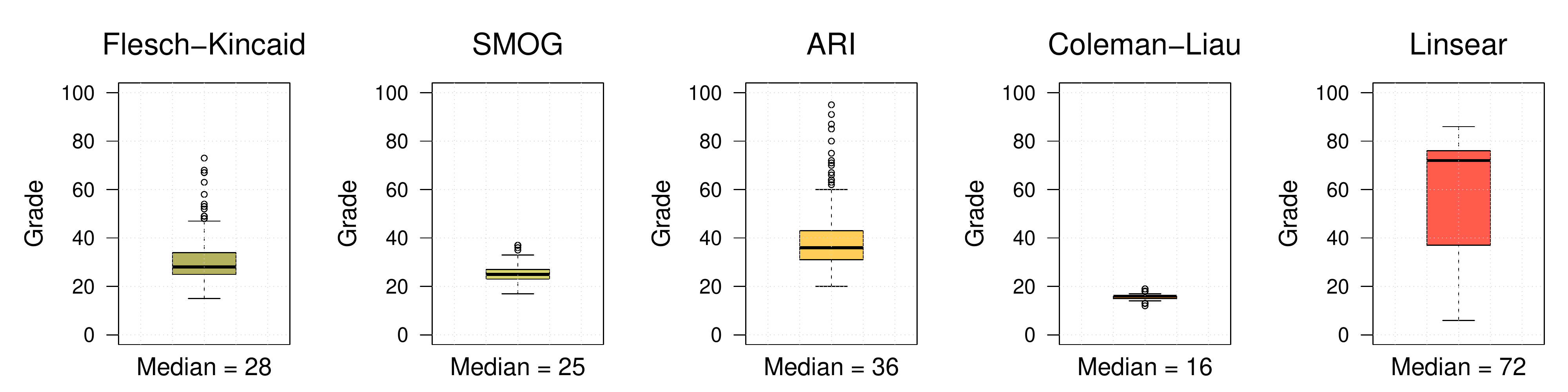}
\caption{Readability Grades ($g_1, \ldots, g_5$, $n = 201$)}
\label{fig: indices}
\end{figure*}

The individual legislations and policy documents are enumerated in
Fig.~\ref{fig: regs}. Of these, about 34\% are decisions, 32\% directives, 18\%
regulations, and the remaining communications, recommendations, and other
document types. The high amount of decisions is partially explained by the
specifications for frequency bands used for electronic and mobile
communication. Based on a subjective classification, as much as 35\% of the laws
and policy documents are about telecommunications. Privacy and data protection
(14\%), copyright and intellectual property (14\%), different contracts and
justice in general (6\%), Internet governance (4\%), and cyber security (3\%)
follow.

\section{Methods}

Five traditional readability indices are used. The selection of these was
practical; these are readily available from an existing Python
package~\cite{textstat21}. All measure the hypothetical grade level required to
comprehend a text. The grades are based on those used in the United States. In
theory, there is a upper limit in the grades---a graduate program in a
university would be somewhere around the 22:th grade or so, but none of the
indices impose limits. Negative values are also possible. Thus, in general, the
higher the score, the more difficult a text is to comprehend. To ensure
comparability, all scores from the indices are further truncated up towards the
nearest integer.

The first is the Flesch–Kincaid index~\cite{Kincaid75}. It is defined as:
\begin{equation*}
g_1 =  \left\lceil 0.39 \left(\frac{\textmd{\#~words}}{\textmd{\# sentences}}\right)
  + 11.8\left(\frac{\textmd{\#~syllables}}{\textmd{\#~words}}\right) - 15.59 \right\rceil,
\end{equation*}
where $g_1$ refers to a grade. The second is the SMOG index or the ``simple
measure of gobbledygook''~\cite{McLaughlin69}. It is given by:
\begin{equation*}
g_2 = \left\lceil 1.0430\sqrt{30\left(\frac{\textmd{\#~polysyllables}}{\textmd{\#~sentences}}\right)} + 3.1291 \right\rceil.
\end{equation*}
The third is ARI, the automated readability index \cite{textstat21, Smith70}:
\begin{equation*}
  g_3 = \left\lceil 4.71
  \left(\frac{\textmd{\#~characters}}{\textmd{\#~words}}\right)
+ 0.5 \left(\frac{\textmd{\#~words}}{\textmd{\#~sentences}}\right) - 21.43 \right\rceil .
\end{equation*}
The fourth is the Coleman-Liau index~\cite{Coleman75}, as defined by:
\begin{equation*}
g_4 = \left\lceil 0.0588\alpha - 0.296\beta - 15.8\right\rceil ,
\end{equation*}
where $\alpha$ is the number of letters per a hundred words and $\beta$ the
average number of sentences per 100 words. The fifth and final index is the
Linsear Write readability formula. It is defined differently from the other four
indices. In essence: for each 100 word sample, easy (with two syllables or less)
and hard (three or more syllables) are counted and scored (with one and three
points, respectively), after which the per-sample scores are divided by the
number of sentences in the sample, and further scaled~(for more details see
\cite{textstat21}). As with the other indices, a truncated grade, $g_5$, is
outputted from the arithmetic.

Two additional points are worth briefly making about these classical
indices. First, there are many modifications to these, as well as numerous
alternatives. Hundreds of individual variables were considered already in the
1970s for constructing readability indices~\cite{Entine78}. More recent
modifications have focused on natural language processing and machine learning
methods~\cite{Francois12, Timana20}. The second point follows: all readability
indices have always been heavily criticized. The criticism and thus limitations
are later on briefly discussed in the concluding Section~\ref{sec:
  conclusion}. For the present purposes, it suffices to justify the use of the
five readability indices with an argument that modifications and more elegant
methods seem uncalled for in the present application domain. As there is no
prior empirical work in the domain, the five indices serve to make a baseline.

\section{Results}

The grades across the $201$ documents are shown in Fig.~\ref{fig: indices} for
each of the five readability indices. There are three points to make from the
figure. First, the Linsear Write index does not pass a commonsense validity
test; the grades from the index show a fairly large variance, but the median of
$72$ is not a realistic value (grade) in practice. Second, there is another
potential validity concern about the Coleman-Liau index; the small standard
deviation of $1.34$ seems surprising when compared to the outputs from the other
four indices. Third, the previous two points translate into modest correlations
with the grades from the remaining three indices. In contrast, the grades from
the Flesch-Kincaid, SMOG, and ARI indices are highly correlated (see
Table~\ref{tab: cor}). By implication, a sum variable constructed from these
three indices (based on the rowwise arithmetic means) attains high internal
consistency. For instance, Cronbach's $\alpha$-coefficient~\cite{Cronbach51} is
as large as~$0.98$.

\begin{table}[th!b]
\centering
\caption{Correlations (Pearson)}
\label{tab: cor}
\begin{tabular}{llrrrrr}
\toprule
& & 1. & 2. & 3. & 4. & 5. \\
\cmidrule{3-7}
1. & Flesch-Kincaid & \\
2. & SMOG & $0.940$ & \\
3. & ARI & $0.995$ & $0.928$ \\
4. & Coleman-Liau & $-0.111$ & $0.018$ & $-0.098$ \\
5. & Linsear & $0.330$ & $0.347$ & $0.312$ & $-0.232$ & \qquad\qquad \\
\bottomrule
\end{tabular}
\end{table}
\begin{figure}[th!b]
\centering
\includegraphics[width=\linewidth, height=5.4cm]{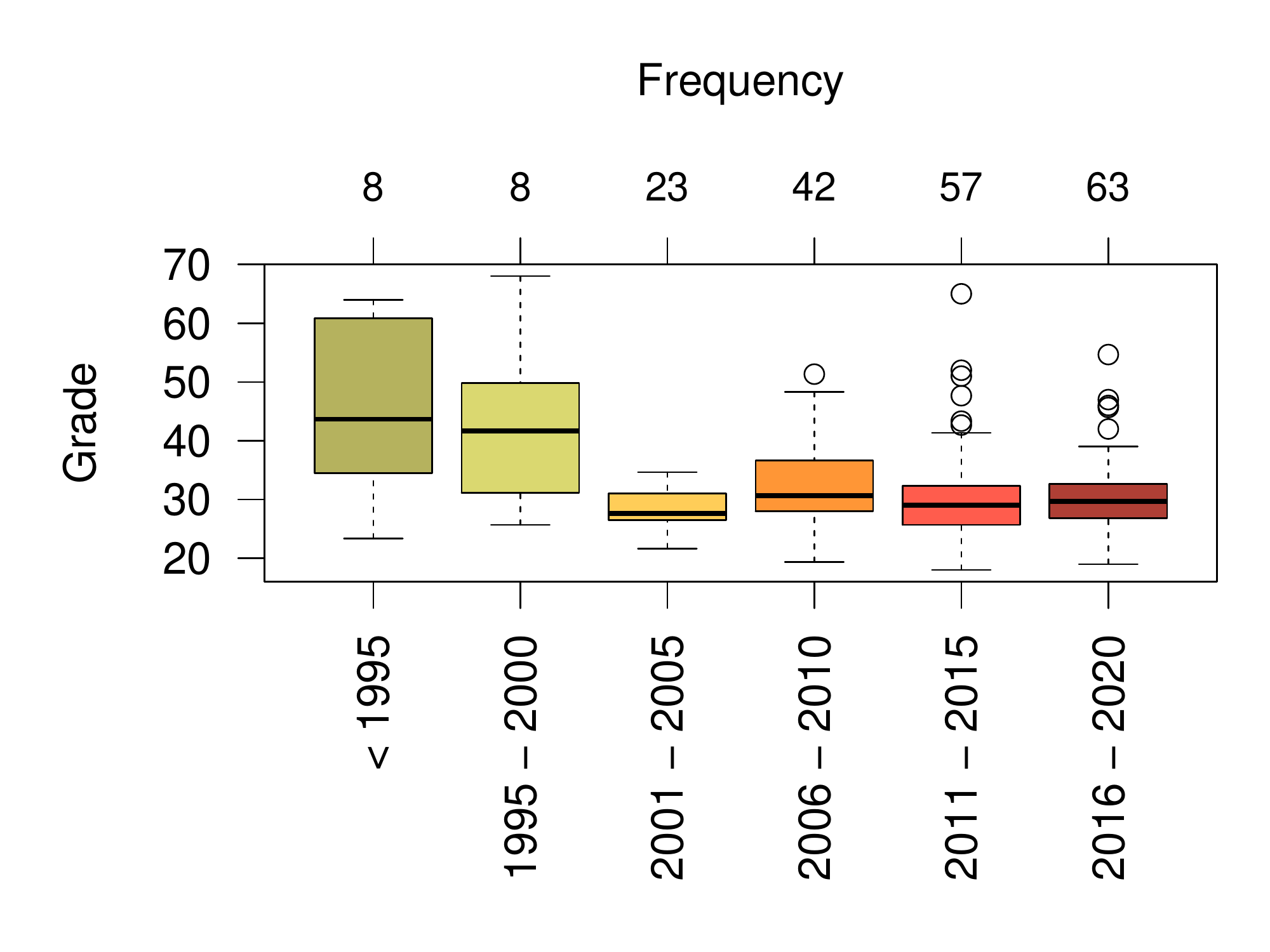}
\caption{Readability Grades Across Time (sum variable)}
\label{fig: years}
\end{figure}

These observations are enough to answer to the first and second research
questions. Regarding RQ$_2$: there are notable differences between the five
indices. Regarding RQ$_1$: based on the median of the sum variable, the overall
readability is somewhere near the $30$:th grade. What does such a grade mean;
how large is this value? Given that even the first quartile is about $27$, it
seems equitable to conclude that a completion of a graduate program is required
to comprehend the DSM laws and policy \text{documents---at} least insofar as the
quantitative readability indices convey their intended function. Even if their
validity is questioned, as can be done on many grounds, there is still a
comparative viewpoint supporting the conclusion of overall complexity. For
instance, local newspapers in the United States have been observed to attain
values around $12$ at maximum~\cite{Wasike16}. Another comparative example would
be abstracts in psychology papers for which values around $17$ have been
reported~\cite{Stricker20}. Finally, the visualization in Fig.~\ref{fig: years}
suffices to answer to RQ$_3$. Although the amount of laws and policy documents
has steadily grown over the years, the readability of these has slightly
improved when compared to the 1980s and 1990s. A potential explanation may
relate to incremental law-making; many of the new laws enacted amend or replace
old laws, building upon their foundations. Nevertheless, the averages have still
remained around $30$ or~so.

\section{Conclusion}\label{sec: conclusion}

This short paper evaluated the readability of laws and policy documents related
to the digital single market of the EU. In general, these are difficult to
comprehend according to the quantitative readability indices. Even though
readability has slightly improved, the hypothetical grade level is still around
thirty. What could explain this result? A partial explanation would be that the
DSM laws and policies, in particular, simply are complex; many of them address
highly technical topics. Another partial explanation stems from the EU's
law-making processes. After a law is finally enacted, it has gone through a
heavy process of revisions, often including copy-pasted snippets from
politicians pressed by lobbyists~\cite{Ruohonen19JPS}. While copy-pasting is not
unique to the EU~\cite{Allee19}, it is likely to increase gobbledygook. Another
source relates to Regulation 1/1958 according to which multiple European
languages must be accounted for already during the law-making processes.

A further potential explanation stems from the limitations of the quantitative
readability indices. In particular, the construct validity of these is often
questionable; it is not entirely clear whether they measure what they intend to
measure. It is not necessary to elaborate the lengthy criticism in detail. It
suffices to note that the indices ignore grammar~\cite{McClure87}, and do not
address the fact that simplicity by itself does not necessarily guarantee
comprehension~\cite{Korunovska20}. Thus, developing a readability index
specifically for law would be a good topic for further research; the newer
machine learning methods may help with the task. Another topic would be to
examine the comprehension of laws and policy documents by human subjects. A
particularly interesting question is the comprehension---or, possibly, lack
thereof---among engineering and computer science students. The question is
relevant because it has been computer science that has been seen to enhance law
and its practice, often with questionable consequences~\cite{Goltz19}, and not
the other way~around.

\bibliographystyle{IEEEtran}


\end{document}